       \newcommand{\nl}{\nonumber \\}
       \newcommand{\beq}{\begin{equation}}
       \newcommand{\eeq}{\end{equation}}
       \newcommand{\beqa}{\begin{eqnarray}}
       \newcommand{\eeqa}{\end{eqnarray}}
       \newcommand{\beqas}{\begin{eqnarray*}}
       \newcommand{\eeqas}{\end{eqnarray*}}
       
       \newcommand{\bnab}{\mbox{\boldmath ${\nabla}$}}

       \newcommand{\x}{\mbox{\boldmath$\times$}}

       \newcommand{\bb}{{\mathbf b}}

       \newcommand{\be}{{\mathbf e}}

       \newcommand{\bv}{{\mathbf v}}
       
       \newcommand{\bx}{{\mathbf x}}

       \newcommand{\bA}{{\mathbf A}}
       \newcommand{\bB}{{\mathbf B}}

       \newcommand{\bE}{{\mathbf E}}

       \newcommand{\bR}{{\mathbf R}}

       \newcommand{\bV}{{\mathbf V}}

       \newcommand{\dadb}[2]{\frac{{  d}#1}{{  d}#2}}

       \newcommand{\paraparb}[2]{\frac{\partial #1}{\partial #2}}

\newcommand{\paral}{\parallel}

\documentclass[12pt]{article}
\usepackage[dvips]{graphicx}\usepackage[dvips]{graphics}
\usepackage{enumerate}
\usepackage{psfrag}
\usepackage{color}
\begin{document}
\begin{center}
{\large \bf
 Vlasov versus reduced kinetic theories for helically symmetric equilibria
 \vspace{2mm}  } \vspace{4mm}

{\large  H. Tasso$^1$, G. N. Throumoulopoulos$^2$} \vspace{2mm}

{\it $^1$Max-Planck-Institut f\"{u}r Plasmaphysik, Euratom
Association,\\
 D-85748 Garching, Germany}
\vspace{2mm}

 { \it  $^2$ University of Ioannina, Association Euratom-Hellenic Republic,\\
 Section of Theoretical Physics, GR 451 10 Ioannina, Greece }
\end{center}
%
%
\vspace{2mm}
\begin{center}
{\bf Abstract}
\end{center}
 \noindent

 A new constant of motion  for helically symmetric equilibria  in the vicinity of the magnetic axis is obtained in the framework of Vlasov theory. In view of this constant of motion the Vlasov theory is compared with drift kinetic and gyrokinetic theories near axis.  It turns out that  as in the case of axisymmetric equilibria [H. Tasso and G. N. Throumoulopoulos, Phys. Plasmas 18, 064507 (2011)] the Vlasov  current density  thereon can differ appreciably from the drift kinetic and gyrokinetic current densities. This indicates  some limitation on the implications of reduced
kinetic theories, in particular as concerns the physics of energetic particles in the central region of
magnetically confined plasmas.
 \vspace{1cm}

 \noindent
{\sf Accepted for publication in Physics of Plasmas}

\newpage
\begin{center}
{\bf \large I.\  Introduction}
\end{center}

Kinetic equilibria may provide  broader  and more precise  information than multifluid or magnetogydrodymanic  equilibria as those governed  by the Grad-Shafranov equation. Since solving self consistently the kinetic equations is tough  particularly in complicated geometries the majority of kinetic equilibrium solutions are restricted to one dimensional configurations in plane geometry, e.g.  \cite{Cha}-\cite{stne}. Of particular interest are equilibria with  sheared toroidal and poloidal flows which play a role in the transition to  improved confinement regimes in tokamaks and stellarators, though  understanding  the physics of this transition  remains incomplete.  Construction of kinetic equilibria is crucially related to the particle constants of motion which the distribution function depends on. In the framework of Maxwell-Vlasov theory only a couple of constants of motions are known for symmetric two-dimensional equilibria, i.e. the energy, $H$,  and the momentum $p_{x3}$ conjugate to the ignorable coordinate, $x_3$,  out of the four potential constants of motion. Therefore, for distribution functions of the form  $f(H,p_{x3})$ only macroscopic flows  and currents  along the  direction associated with $x_3$ can be derived, e.g. toroidal flows for axisymmetric plasmas.  The creation of poloidal flows requires  additional constant(s) of motion. This  remains an open question despite the fact that in a previous study    \cite{TaTh} we found locally in the vicinity of the magnetic axis  the following new constant of motion:
 $C=v_z+I \ln \left| v_\phi\right|$, where $v_\phi$ is the toroidal particle velocity, $v_z$ the  velocity component parallel to
 the axis of symmetry and $\bB_\phi = I/r \be_\phi$  the toroidal magnetic field near axis ($r, \phi, z$ are cylindrical
 coordinates). An additional  open question remains  a potential extension of the proof of non existence of   magnetohydrodynamic axisymmetric equilibria with purely poloidal flows  \cite{ThWe} to Vlasov equilibria. A third constant of motion was studied extensively in the Astrophysics literature \cite{st}-\cite{efvo}.   In particular for axisymmetric astrophysical systems distribution functions depending on an approximate  third constant of motion result in velocity ellipsoids with unequal axes in agreement with observational data for our Galaxy \cite{co,co1,efvo}.  

Owing to the contemporary  and probably  future limited computational efficiency for simulations in the framework  of Vlasov theory  approximate  kinetic theories have been established in reduced phase spaces   as the drift kinetic and gyrokinetic ones.  In
 both theories the reduced phase space is five dimensional with three spatial components associated with the guiding center
 position, $\bf R$ (or the gyrocenter position in the framework of gyrokinetic theory),  and a velocity component  parallel to the  magnetic field, $v_\paral$;
  also, the two components  of the
 perpendicular particle  velocity are approximated after a gyroangle averaging with the magnetic moment which is treated as an adiabatic invariant. A related underlying assumption  for both  reduced theories is that the ratio $\epsilon$ of the gyroperiod
 to the macroscopic time scale is small.
 In the drift kinetic theory $\epsilon$ is the same as the ratio of the gyroradius to macroscopic scale length while in
 the gyrokinetic theory small spatial variations are permitted.  Because of the reduction of the phase space some information of the particle motion is missed.
In this respect it may be noted  that the
reduced-phase-space kinetic theories are developed via
expansions in $\epsilon$, the convergence of which is not guaranteed. This gives rise to the question: is the missing information important? To address this question we compared  axisymmetric Vlasov equilibria near the  magnetic axis with drift kinetic and gyrokinetic ones \cite{tath1}. It turned out that because of missing  the above constant of motion, C,  in the latter case  the on axis current density  can differ appreciably from the ``actual" Vlasov current density. Also, unlikely  Vlasov theory, the reduced kinetic  theories can not distinguish a straight from a  curved magnetic axis. 

The aim of the present contribution is twofold: first to extend the above local constant of motion, $C$, to the more general class of helical symmetric equilibria and second to compare the Vlasov helical symmetric equilibrium   characteristics near the magnetic axis with respective drift kinetic and gyrokinetic characteristics. The new local constant of motion is derived in Sec. II. Then, Vlasov theory is compared near axis with reduced kinetic theories in Sec. III. Sec. IV summarizes the conclusions. 

\begin{center}
{\bf \large II.\ A third   Vlasov constant  of motion near magnetic axis}
\end{center}
We will employ the following form of Vlasov equation \cite{sata}:
\begin{equation}
\frac{\partial f}{\partial t} + {\bf v}\cdot  \bnab f + {\bf e}_{i}\cdot
({\bf E} + {\bf v}\times {\bf B})\frac{\partial f}{\partial v_{i}} +
({\bf e}_{i}\cdot {\bf v}\times  \bnab \times {\bf v})\frac{\partial f}{\partial
v_{i}} = 0.
                                                \label{vlas}
\end{equation}
To derive  (\ref{vlas}) one  uses  general orthogonal coordinates  $(x_1,x_2, x_3)$ with unit basis vectors $\be_i= \bnab x_i/| \bnab x_i|$ ($i=1,2,3$), expresses the ``microscopic fluid" velocity in the basis of $\be_i$ as $\bv=v_i \be_i$ and uses the ``microscopic fluid" momentum equation, 
$$
\frac{\partial {\bf v}}{\partial t} + {\bf v}\cdot \bnab {\bf v} = {\bf E} +
{\bf v}\times {\bf B},
$$
where ${\bf E}$ and ${\bf B}$ are the electric and magnetic fields
consistent with Maxwell equations. 
The term ``microscopic fluid" relates to the fact that the Vlasov equation is an approximation to the 
N-particles Liouville equation, which replaces the N particles by
a continuum. This is sometimes termed ``fluid approximation". To avoid confusion, however,  the usual term ``particle" will be adopted in place of ``microscopic fluid".  Also, for the sake of notation simplicity and without loss of generality we will consider only ions
   and employ convenient units
  by setting $m=q=c=1$ where $m$ and $q$ are
 the ion mass and charge and $c$ is the velocity of light.

In connection with the helical symmetry  
we specify the coordinates as ($x_1=r$, $x_2=a\phi+\beta  z$, $x_3=-\beta \phi+a z$), where $a$ and $\beta$ are parameters. Helical symmetry means that any quantity does not depend on $x_3$. Translational symmetry and axisymmetry are then recovered as particular cases for ($a=1$, $\beta=0$) and ($a=0$, $\beta=1$).  The respective unit vectors are
$$
\be_1=\be_r,\ \  \be_2=\frac{a\be_\phi+\beta r \be_z}{\left(a^2+\beta^2 r^2\right )^{1/2}}, \ \  \be_3=\frac{ar \be_z-\beta  \be_\phi}{\left(\beta^2+a^2 r^2\right)^{1/2}}.
$$
Note that these basis vectors are in general non orthogonal.  A helical magnetic axis however is located at a constant distance $r$ from the $z$-axis. Since we will study the equilibrium near the magnetic axis we set  $r=1$ thereon and make the choice $a^2+\beta^2=1$. Consequently,   the above basis vectors reduce to  ($\be_1=\be_r$, $\be_2=a\be_\phi+\beta \be_z$,  $\be_3=-\beta \be_\phi+a \be_z$) and  become orthogonal on axis,  with $\be_2$ along and $\be_3$ perpendicular to the axis. Thus,   (\ref{vlas}) can be applied in the neighbourhood  of  axis.  The magnetic field near axis can be written in the form
$$
\bB=\frac{I}{r}\be_\phi+B_0 \be_z=\frac{I}{r}\frac{a \be_2-\beta\be_3}{a^2+\beta^2}
     +B_0\frac{\beta \be_2+a \be_3}{a^2+\beta^2}.
$$
By the choice $a B_0=\beta I$ we take on axis $\bB=I/a \be_2$ as it should be. Also, we examine the case of $\bE=0$ on axis but $ \bnab f \neq 0$ thereon. The latter assumption  may be regarded as extraordinary because $ \bnab f$ is related to the density gradient on axis which for usual peaked density profiles  is expected to vanish.   The reason for assuming $ \bnab f \neq 0$ on axis is that  for axisymmetric equilibria,   if   (\ref{vlas}) is solved near axis by the method of characteristics under the assumption $ \bnab f=0$ on axis  the usual  toroidal momentum constant of motion, $p_\phi$, is missed,   while $p_\phi$ is recovered if $ \bnab f \neq 0$ thereon \cite{tath1}. 

 Using the above basis for the helically symmetric equilibria under consideration Eq. (\ref{vlas}) near axis becomes 
\beq
v_1\paraparb{f}{x_1}+v_2\paraparb{f}{x_2}+w_1\paraparb{f}{v_1}
+w_2\paraparb{f}{v_2}+w_3\paraparb{f}{v_3}=0,
                                                  \label{vlas1}
\eeq
where
\beq
w_1=\left(a v_2-\beta v_3\right)^2-v_3\frac{I}{a}, \ \ w_2=a v_1 \left(\beta v_3-a v_2\right), \ \  w_3=v_1\frac{I}{a} + v_1 \beta\left(a v_2-\beta v_3\right).
                                                    \label{www}
\eeq 
The characteristics of (\ref{vlas1}) are given by the solutions of 
\beq
\dadb{x_1}{v_1}=\dadb{x_2}{v_2}=\dadb{v_1}{w_1}=\dadb{v_2}{w_2}=\dadb{v_3}{w_3}. 
                                                     \label{char}
\eeq
Integration of the last equation yields the new constant of motion:
\beqa
C&=&v_3+\frac{\beta(\beta^2-a^2)(\beta v_3-a v_2)+a I \ln \left| \left\lbrack a^4 v_2-a^2 
  \beta^2 v_2 +a \beta^3 v_3 -\beta(I+a^3 v_3)\right\rbrack \right| }{\left(a^2-\beta^2\right)^2}
\label{3const}  \nl
& &  \ \ \mbox{for} \ \  a \neq \beta, \\
C&=& 2 v_3-v_2-\frac{a^3}{I}\frac{\left(v_2-v_3\right)^2}{2} \ \ \mbox{for} \ \ a=\beta. \label{3const1}
\eeqa
The respective axisymmetric constant of motion, $C=v_z+I\ln|v_\phi| $, is recovered from (\ref{3const}) for $(a=1, \beta=0)$ and the translational symmetric constant of motion, $C=v_z$,   for $(a=0, \beta=1)$. Also,  conservation of energy $H=1/2(v_1^2+v_2^2+v_3^2)$ follows from the manipulation of equations (\ref{char}) presented in Appendix. Distribution functions of the form $f(H, C)$ could create a helical current on axis because of the  dependence of $C$ on $v_2$ additionally to that in the logarithmic term. This is different from the axisymmetric case in which  toroidal axisymmetric currents on axis are not possible at all for the same class of distribution functions because of the  dependence of $C$ on $|v_\phi|$. This difference may be expected because for translational symmetric equilibria which are recoverable from the helically symmetric ones the respective distribution functions $f(H,C)=f(H,v_z)$ can create currents on axis. However, such a helical equilibrium is not physically acceptable  because of the dependence of $C$ on $v_3$ (but not on $v_1$)  implying the  creation of a flow perpendicular to the axis. Similar unphysical flows are created in the axisymmetric case because of the respective dependence of $C$ on $v_z$ (but not on $v_r=v_1$). As discussed in \cite{TaTh} possible reasons of this unphysical behaviour are: (i) the lack of a potential fourth constant of motion involving  $v_1$  in such a way that  creation of  regular poloidal flows  be possible, (ii) non taking into account here  the MHD property of coincidence of the magnetic surfaces with the flow surfaces, (iii) additional drifts because of the curvature and torsion of the magnetic field which are eliminated in translational symmetric geometry and (iv) potential damping of unphysical flows in the framework of a collisional kinetic theory. The above consideration  shows that straight, circular and helical magnetic axes are well distinguished  in the framework of Vlasov theory.

Unlikely  the axisymmetric case \cite{tath1} the  generalized momentum constant of motion, $p_{x3}= v_3+A_3$, where $\bA(x_1,x_2)$ is the vector potential,   can not follow  from (\ref{char}) though we have assumed $\bnab f \neq 0$ on axis. This is obtained form the Lagrangian  
$$
L=\frac{1}{2}(\dot{x}_1^2+ + \dot{x}_2^2 + \dot{x}_3^2) + \Phi(x_1, x_2) + \dot{\bx} \cdot{\bA}(x_1,x_2)
$$ even in the presence of an electric filed on axis associated with the electrostatic potential $\Phi$. Also, a generic non local  derivation off $p_{x3}$ is provided in \cite{copf}.   

\begin{center}
{\bf \large III.\  Comparison of Vlasov with drift kinetic and gyrokinetic theories.}
\end{center}

The near axis consideration of helically symmetric equilibria will be repeated  in the framework of the drift kinetic and gyrokinetic theories which were employed in Ref. \cite{tath1} on an individual basis below. For convenience a brief review of these theories  will also   be given  here because otherwise frequent reference to \cite{tath1} would make reading of the present section tedious.   \newline

\noindent
 {\large \em Drift kinetic theory }
 \newline

The drift kinetic theory established in Ref. \cite{CoPf} is based on
 the Littlejohn's Lagrangian for the guiding center  motion \cite{Li}   extended
 to include the polarization drift in such a way that local
 conservation of energy is guaranteed. Introducing the modified potentials, $\bA^\star$ and
 $\Phi^\star$,  this Lagrangian can be written in the concise form \cite{PfMo}
\beq
L=\bA^\star\cdot\dot{\bR}-\Phi^\star,
                                         \label{little}
\eeq
where
 \beqa
 \bA^\star&=& \bA +v_\paral \bb +\bV_E,  \label{2a} \\
 \Phi^\star&=& \Phi + \mu B + \frac{1}{2}(v_\paral^2+v^2_E),  \label{2b}\\
 \bV_E&=&\frac{\bE \x \bB}{B^2}.
 \eeqa
Here,    $\Phi$ and $\bA$ are the usual electromagnetic scalar and vector  potentials and
 $\bb=\bB/B$.
 The drift kinetic
 equation for the guiding center distribution function $f(\bR, v_\paral, \mu, t)$ (with
 $\dot{\mu}=0$)
 acquires   the form
 \beq
 \paraparb{f}{t}+ \bV\cdot\bnab f +\dot{v}_\paral
 \paraparb{f}{v_\paral}=0.
                                                   \label{1}
 \eeq
 By means  of the Euler-Lagrange equations following from (\ref{little}) the guiding center velocity, $\bV$, and the ``acceleration" parallel to the magnetic
 field,
 $\dot{v}_\paral$,  can be expressed 
as:
 \beq
\bV=\dot{\bR}=v_\paral\frac{\bB^\star}{B_\paral^\star}+\frac{\bE^\star\x\bb}{B_\paral^\star},
                                                                 \label{3}
 \eeq
  \beq
  \dot{v}_\paral= \frac{\bE^\star\cdot\bB^\star}{B_\paral^\star}=\frac{1}{v_\paral}\bV\cdot\bE^\star,
                                                           \label{4}
  \eeq
  where
\beq
 \bE^\star= -\paraparb{\bA^\star}{t}-\paraparb{\Phi^\star}{\bR}, \ \ 
\bB^\star=\bnab\x \bA^\star, 
\eeq
  \beq
  B_\paral^\star=\bB^\star\cdot \bb=B +v_\paral\bb\cdot \bnab \x\bb+ \bb\cdot
  \bnab\x\bV_E.
                                                         \label{5}
  \eeq
  Explicit expressions for $\bV$ and $\dot{v}_\paral$ are given by Eqs. (3.24) and (4.17)
  of Ref. \cite{CoPf}. Also, it is noted here that Eqs. (\ref{3})
  and (\ref{4}) have similar structure as the respective gyrokinetic
  equations of Refs. \cite{PfCo1} and \cite{CoPf1} [Eqs. (5.39) and
  (5.41) therein].
  Since $B_\paral^\star$ appears  in the denominators of (\ref{3})
  and (\ref{4}) a singularity occurs for $B_\paral^\star=0$. For $\bE=0$ this singularity can be expressed by the
  critical parallel velocity $v_c=-\Omega/(\bb\cdot \bnab \x \bb)$, where $\Omega$ is the gyrofrequency.
  Therefore,  the theory is singular for large $|v_\paral|$ at which
  $\bV$ and $\dot{v}_\paral$ diverge and consequently non-casual
  guiding center orbits occur and the guiding center conservation in
  phase space is violated \cite{CoWi}. It is the
  $v_\paral$-dependence of $\bA^\star$ [Eq. (\ref{2a})] that
  produces the singularity. In order to regularize the singularity
  $v_\paral$ in (\ref{2a}) can be replaced by an antisymmetric
  function $g(z)$ with $z=v_\paral/v_0$, where $v_0$ is some constant
  velocity \cite{CoWi,CoPf,PfMo}. The nonregularized theory presented here for simplicity
  is obtained for $g(z)=z$. In the regularized theory $g(z)\approx
  z$ should still hold for small $|z|$. For large $|z|$, however, $g$
  must stay finite such that with $v_0 \gg v_{\mbox{\small thermal}}$ one has $v_0
  g(\infty) < v_c$. A possible  choice for $g(z)$ is $g(z)=\tanh z$.

Since for  a helically symmetric  equilibrium   the magnetic field on axis becomes purely helical and dependent
 only on $r$, 
 one readily calculates near axis with $\bE=0$ thereon:

 \beq
\bnab \x \bb=\bnab \x \be_2 =\frac{a}{r} \be_z, \ \ \bB^\star=\bB+v_\paral \frac{a}{r} \be_z,\ \ B^\star_\paral = B+v_\paral\frac{a}{r}\beta
                                                         \label{hel1}
\eeq
\beq
 \bnab
 B(r)=\frac{dB}{dr}\be_r, \ \  \Phi^\star=\mu B+ \frac{1}{2} v_\paral^2, \ \ \bE^\star=-\mu\dadb{B}{r}\be_r
                                                     \label{hel2}
\eeq
  and consequently
 \beq
 \bV=v_\paral \frac{B}{{B^\star}_\paral}
 \be_2+ \frac{v_\paral^2 a}{r B^\star_\paral }\be_z-\frac{\mu}{B^\star_\paral}\dadb{B}{r}\be_3,
                                                  \label{6}
 \eeq
 \beq
  \dot{v}_\paral=0.
                                                 \label{7}
  \eeq
  As expected on axis the guiding center
 velocity consists of a component parallel to $\bB$ and the curvature
 and grad-$B$ drifts.  
 Also,  the ``acceleration" $\dot{v}_\paral$ vanishes because there is  no
 parallel force and  the drift kinetic equation (\ref{1}) becomes
\beq \bV\cdot\paraparb{f}{\bR}=0.
                                                \label{7a}
                                                  \eeq
Because of (\ref{7}) $v_\paral$ is a constant of motion and therefore distribution functions of the form $f(H, v_\paral)$, where $H=\mu B+1/2 v_\paral^2$, either can or can't produce helical currents by choosing $f$ either odd or even function of $v_\paral$. This property holds also for axisymmetric and translational symmetric equilibria \cite{tath1} because (\ref{7}) keeps valid irrespective of the kind of symmetry. Thus, unlike Vlasov theory the drift kinetic theory can not distinguish equilibria of straight, circular or   helical magnetic axes. Also, according to the results of section III the respective obtainable near axis Vlasov class of distribution functions $f(H,C)$ can not produce physically acceptable currents on axis\footnote{Physically acceptable Vlasov currents can be produced  by the class $f(H, p_{x3})$ which, however, corresponds to   drift kinetic distribution functions of the form $f(H, p_{x3}^d)$, where near axis $p_{x3}^d=A_3+v_\paral b_3$, as it follows from the Lagrangian (\ref{little}).}.  
Consequently, since near axis the overwhelming majority of the particles
 are passing, the parallel currents constructed  in the framework of the drift kinetic theory
 may
 differ
 from the ``actual"  ones. Also, unlike in the case of axisymmetric equilibria \cite{tath1},  the $B_\paral^\star$-singularity is present as indicated by Eq. (\ref{hel1}). 
\newline

 \noindent
 {\large \em Gyrokinetic theory }
 \newline

 We will use the gyrokinetic equations of Ref. \cite{Ha} which
 have been employed to a variety of applications (see for example   Refs.
 \cite{LaMc,FeLi,ZhLi}). Eq. (\ref{1}) remains identical in form
 where
 $f(\bR, v_\parallel, \mu, t)$ is now the gyrocenter distribution function
 for ions. The gyrocenter velocity and ``acceleration" are given by
 \beq
 \bV=v_\paral \bb_0
 +\frac{B_0}{B_{0\paral}^\star}\left(\bV_E+\bV_{ \bnab
 B}+\bV_c\right),
                                                    \label{8}
 \eeq
 \beq
 \dot{v}_\paral=-\frac{1}{v_\paral}\bV\cdot\left(\bnab
 \overline{\Phi}+\mu \bnab B_0\right).
                                                    \label{9}
 \eeq
 Here, $\bB_0$ is the equilibrium magnetic field, $\bb_0=\bB_0/B_0$,
 $$B_{0\paral}^\star=(\bB_0+v_\paral\bnab\x\bb_0)\cdot \bb_0,$$
 $ \overline{\Phi}$ stands for the perturbed gyroaveraged
 electrostatic potential, and the  $\bE\x\bB$-drift velocity $\bV_E$, the grad-$B$ drift velocity
 $\bV_{ \bnab B}$,  and the curvature drift velocity $\bV_c$ are given
 by
 \beq
 \bV_E=-\frac{\bnab \overline{\Phi}\x\bnab B_0}{B_0^2},
                                                    \label{10}
 \eeq
 \beq
 \bV_{ \bnab B}=\frac{\mu}{\Omega B_0}\bB_0\x\bnab B_0,
                                                   \label{11}
 \eeq
 \beq
 \bV_c=\frac{\mu v_\paral^2}{\Omega B_0^2}
 \bb_0\x\bnab\left(p_0+\frac{B_0^2}{2}\right).
                                                \label{12}
 \eeq
 Note that as in the case of  drift kinetic theory a similar singularity occurs at
 $B_{0\paral}^\star=0$. In numerical applications this singularity was
 ``avoided"    by approximating $B_{0\paral}^\star=B_0$ (see for example  Refs.   \cite{LaMc,FeLi}).
 Consideration of the above equations  for a helically symmetric equilibrium   with $\bE=0$ near axis
 yields  relations similar  to (\ref{6}), (\ref{7}) and (\ref{7a}). Therefore,  the  above
 found
  discrepancies of the drift kinetic theory
  with the Vlasov one persist in the framework of the gyrokinetic theory.
  The structure of the reduced kinetic equations in conjunction with
  the symmetry of the equilibrium considered  clearly indicate that
  this conclusion is independent of the particular drift kinetic or
  gyrokinetic equations.

  \begin{center}
{\bf \large IV.\  Summary}
\end{center}

We have found a new constant of motion near the magnetic axis of helically symmetric equilibria in the framework of Vlasov theory [Eqs. (\ref{3const}) and (\ref{3const1})]. 
On account of this constant of motion   a comparison of the Vlasov equation with either the
  drift kinetic or the gyrokinetic equation indicates that the current densities near the magnetic axis in the former case may be different from the ones   in the latter  case. Also, unlike Vlasov,  the reduced kinetic theories can not   distinguish equilibria  with
  straight, circular or helical magnetic axes. 
   This  discrepancy is due  to  the loss of  new  
  Vlasov constant of motion  in the reduced phase space thus indicating 
   that this reduction,  even  if  made rigorously so that local conservation laws and Liouvillean
  invariance of the volume element
   is guarantied, is associated with the loss of  nontrivial physics. This  could put certain limits on  the conclusions from drift kinetic or gyrokinetic
   simulations for time scales large compared to the gyroperiod which is the case of an equilibrium. Though the derivations of both theories are formally correct the convergence of the $\epsilon$-expansions is presumably not uniform for all times which could lead to the noticed discrepancies calculated in this study.   

In addition, a singularity
  which occurs in both drift kinetic and gyrokinetic theories  for large parallel
  particle velocities, present for helically symmetric equilibria,  is usually eliminated in the literature by a rough
  approximation.  Alternative to the rather artificially imposed  regularization reported in Sec. III,  one might  remove this singularity in the integrals associated  with  moments of the drift kinetic equations (see the generic relations (8.14)-(8.17) of \cite{CoPf} for the self consistent charge, current, energy and energy flux densities) by Cauchy principle value integration provided that such an integration would be physically justifiable. This requires further investigation.  

 \begin{center}
 {\large \bf Appendix:\ \ Energy conservation near axis}
 \end{center}

 \renewcommand{\theequation}{{A}\arabic{equation}}
 \setcounter{equation}{0}

From 
$$\frac{dx_1}{v_1}=\frac{dv_1}{w_1} \ \ \mbox{and}  \  \ \frac{dx_2}{v_2}=\frac{dv_2}{w_2} $$
[see Eqs.  (\ref{char})] it follows
\beq
w_1 dx_1 + w_2 d x_2= \frac{1}{2} d(v_1^2+v_2^2) 
                                 \label{a1}
\eeq
Also, employing  
$$\frac{dx_1}{v_1}=\frac{dx_2}{v_2}=\frac{dv_3}{w_3}$$
one has
\beq
w_1 dx_1 + w_2 d x_2=(w_1v_1+w_2 v_2)\frac{dv_3}{w_3}
                                           \label{a2}
\eeq
Explicit calculation  by means (\ref{www}) 
yields 
\beq
w_1v_1+w_2 v_2=-w_3v_3.
                                         \label{a3}
\eeq
From (\ref{a1}), (\ref{a2}) and (\ref{a3}) it follows
$$
\frac{1}{2} (v_1^2+v_2^2+v_3^2)= \mbox{const.}
$$

 \section*{Acknowledgements}


 Part of this work was conducted during a visit of  G.N.T.
 to the Max-Planck-Institut f\"{u}r Plasmaphysik, Garching.
 The hospitality of that Institute is greatly appreciated.

 This work was performed  within the participation of the University
of Ioannina in the Association Euratom-Hellenic Republic, which is
supported in part by the European Union and by the General
Secretariat of Research and Technology of Greece. The views and
opinions expressed herein do not necessarily reflect those of the
European Commission.
\newpage

\vspace*{-1.5cm}

 \end{document}